\documentclass{ws-ijmpcs}

\begin{document}

\markboth{Jakob Salfeld-Nebgen}
{Search for the Higgs Boson decaying into tau pairs}

\catchline{}{}{}{}{}
%
%%%%%%%%%%%%%%%%%%%%%%%%%%%%%%%%%%%%%%%%%%%%%%%%%%%%%%%%%%%%%%%%%%%%

\title{SEARCH FOR THE HIGGS BOSON DECAYING INTO TAU PAIRS}

\author{Jakob Salfeld-Nebgen}

\address{Deutsches Elektronen-Synchrotron, Notkestrasse 85,  22607 Hamburg, Germany\\
jakob.salfeld@cern.ch}

\maketitle

\begin{history}
\received{Day Month Year}
\revised{Day Month Year}
\end{history}

\begin{abstract}
A search for the Standard Model Higgs Boson decaying into $\tau$ pairs is performed using events recorded by the CMS experiment at the
LHC in 2011 and 2012.  
An excess of events is observed over a broad range of Higgs mass hypotheses, with a maximum local significance of 2.93 standard deviations 
at $m_H= 120$ $\textrm{GeV/c}^{2}$. The excess is compatible with the presence of a standard-model Higgs boson of mass 125 $\textrm{GeV/c}^{2}$.
\keywords{}
\end{abstract}

\ccode{PACS numbers:}

\section{Introduction}	
We present a search for the Standard Model Higgs Boson decaying into tau-pairs in the invariant mass region 110-145 $\textrm{GeV/c}^{2}$ performed using events 
recorded by the CMS experiment at the LHC in 2011 and 2012 corresponding to an integrated luminosity of 4.9 fb$^{-1}$ at a centre-of-mass energy of 7 TeV and 
19.4 fb$^{-1}$ at 8 TeV~\refcite{TauPas}. Since July 4th 2012 an experimental observation of a new Boson of a mass around 125 $\textrm{GeV/c}^{2}$ and compatible with the Standard Model Higgs Boson hypothesis is established both by 
the Atlas and CMS collaborations Ref.~\refcite{CMSHiggs}, ~\refcite{AtlasHiggs}, ~\refcite{CMSHiggsSM}. The observation of the Higgs-like Boson is predominantly 
based on its couplings to gauge bosons. To seek direct evidence for its fermionic couplings, the search for the Standard Model Higgs Boson decaying into tau-pairs is a distinguished channel to be
analyzed with the CMS detector. 

\section{Analysis Strategy}
\begin{figure}[pb]
\centerline{\includegraphics[width=4.7cm]{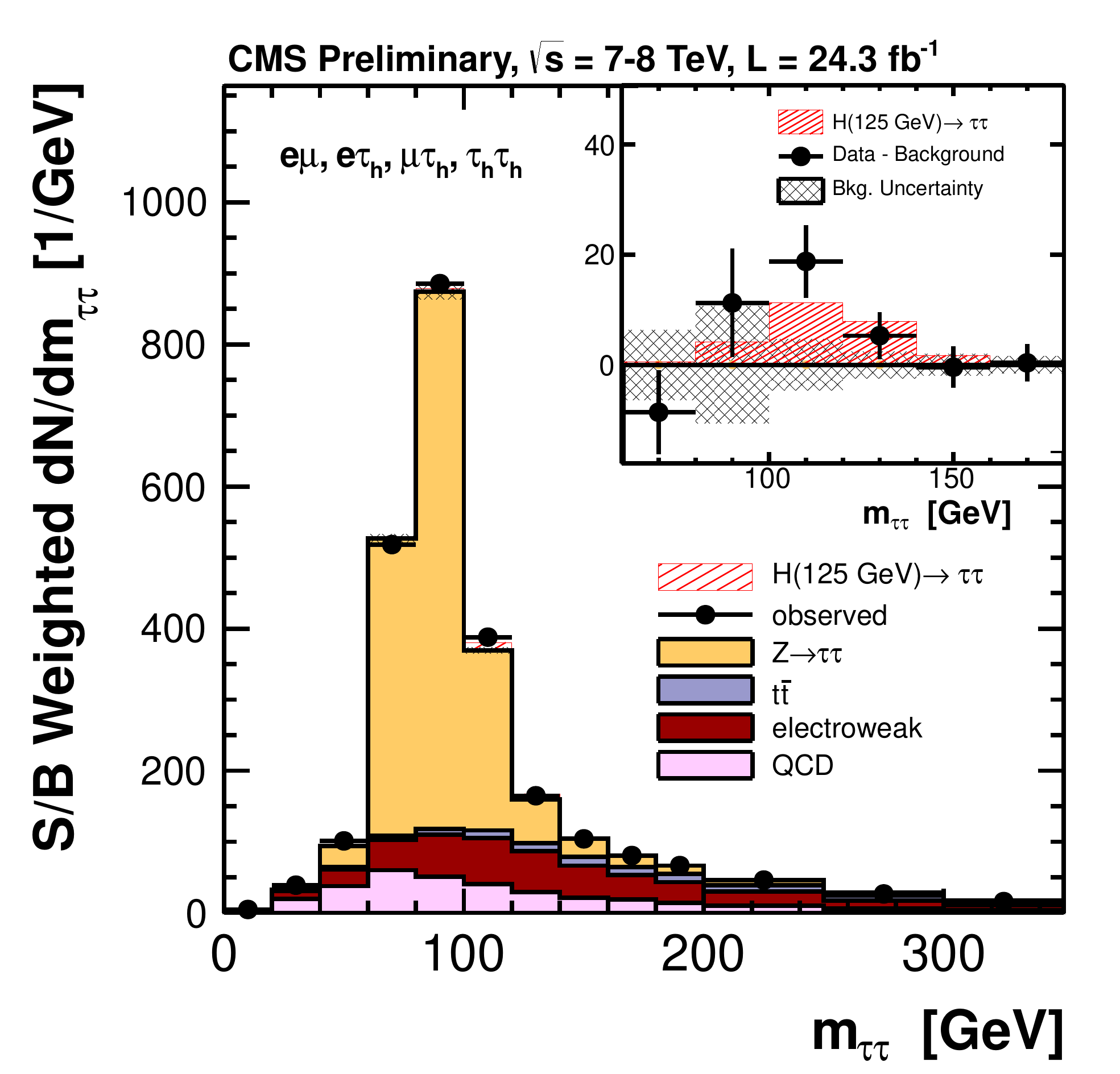}
\hspace{0.1cm}
\includegraphics[width=4.7cm]{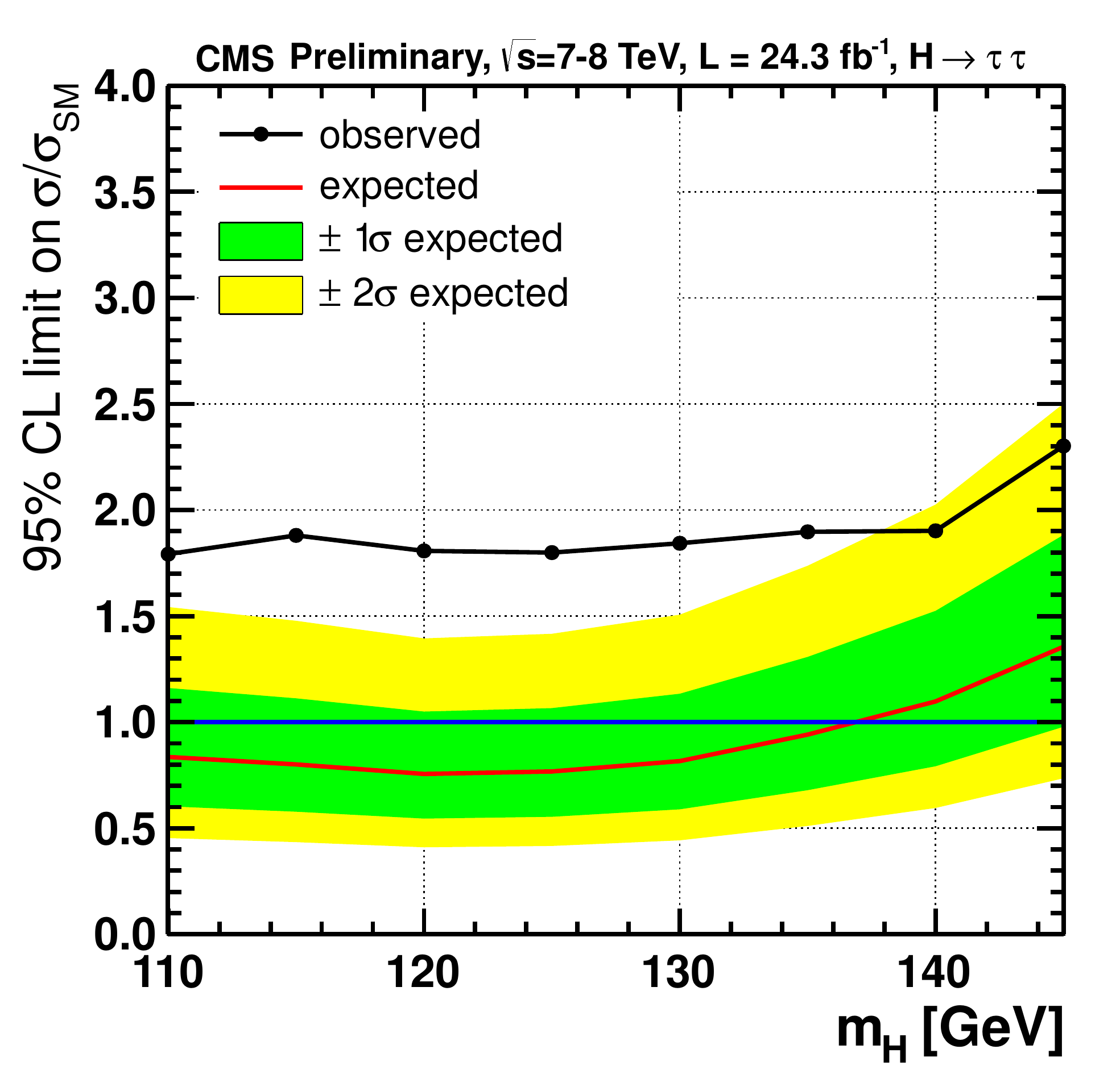}}
\vspace*{8pt}
\caption{ Left: Combined $m_{\tau\tau}$ distribution of $\tau_{h}\tau_{h}$, $e\tau_{h}$, $\mu\tau_{h}$ and $e\mu$ final states weighted 
by the ratio of the expected signal and background yields in each category. Right: 
Observed and expected 95\% CL exclusion limit on the Standard Model Higgs Boson production rate\label{f3}}
\vspace*{0.5in}
\end{figure}

The search is simultaneously performed in several final states of the decaying di-tau system, henceforth denoted as: 
$\tau_{h}\tau_{h}$, $e\tau_{h}$, $\mu\tau_{h}$, $e\mu$ and $\mu\mu$, where $\tau_{h}$ declares the hadronically decaying tau final states. 
For each final state the events are then divided into several disjoint event categories in order to enhance the overall sensetivity by exploiting the 
distinct event topologies of the Higgs production processes. Major Higgs production mechanisms considered are processes via gluon fusion, 
vector boson fusion and the vector boson associated production. The dedicated $H\rightarrow\tau\tau$ analyses exploiting the associated Higgs 
productions (see Ref.~\refcite{CMSTauAssoPas}) are not discussed in detail in this review, however are included in the results in figure ~\ref{f1} and ~\ref{f3}. 

The most sensitive category is the Vector Boson Fusion (VBF) category where the event topology of Standard Model Higgs production via Vector Boson Fusion 
is exploited which is characterized by two high $p_{T}$ jets in the forward regions of the detector. In addition to the event and lepton selection criteria 
events in the VBF category are required to have two jets with $p_{T}>30 \textrm{ GeV/c}$ and $|\eta|>4.7$, having an invariant mass of 
$m_{jj}>500 \textrm{ GeV/c}^{2}$, with a seperation in the pseudo-rapdity $\Delta\eta_{jj} > 3.5$ and no addtional jet with $p_{T}>30 \textrm{ GeV/c}$ in the eta gap between the two corresponding jets.

In addition a 1-Jet category is defined. Events with at least one jet with $p_{T}>30 \textrm{ GeV/c}$ and $|\eta|>4.7$ are selected in this category, 
exluding events of the VBF category. The Standard Model Higgs production via gluon fusion dominates the signal contribution in the 1-Jet category.

Further the 0-Jet category is defined for events with no jet with $p_{T}>30 \textrm{ GeV/c}$ and $|\eta|>4.7$. This category is largely dominated by background processes
and is used to constrain the backgrounds in the VBF and 1-Jet categories.

For all categories a b-tagged jet with $p_{T}>20 \textrm{ GeV/c}$ veto is applied to decouple the analysis from the MSSM Higgs search in the di-tau channel.

% For the $\mu\mu$ final state the overwhelming $Z/\gamma*\rightarrow \mu\mu$ background is reduced with a selection criteria on dedicated multivariate Boosted Decision Trees
% in each category.
% (see figure~\ref{f2}).

\begin{figure}[pb]
\centerline{\includegraphics[width=4.7cm]{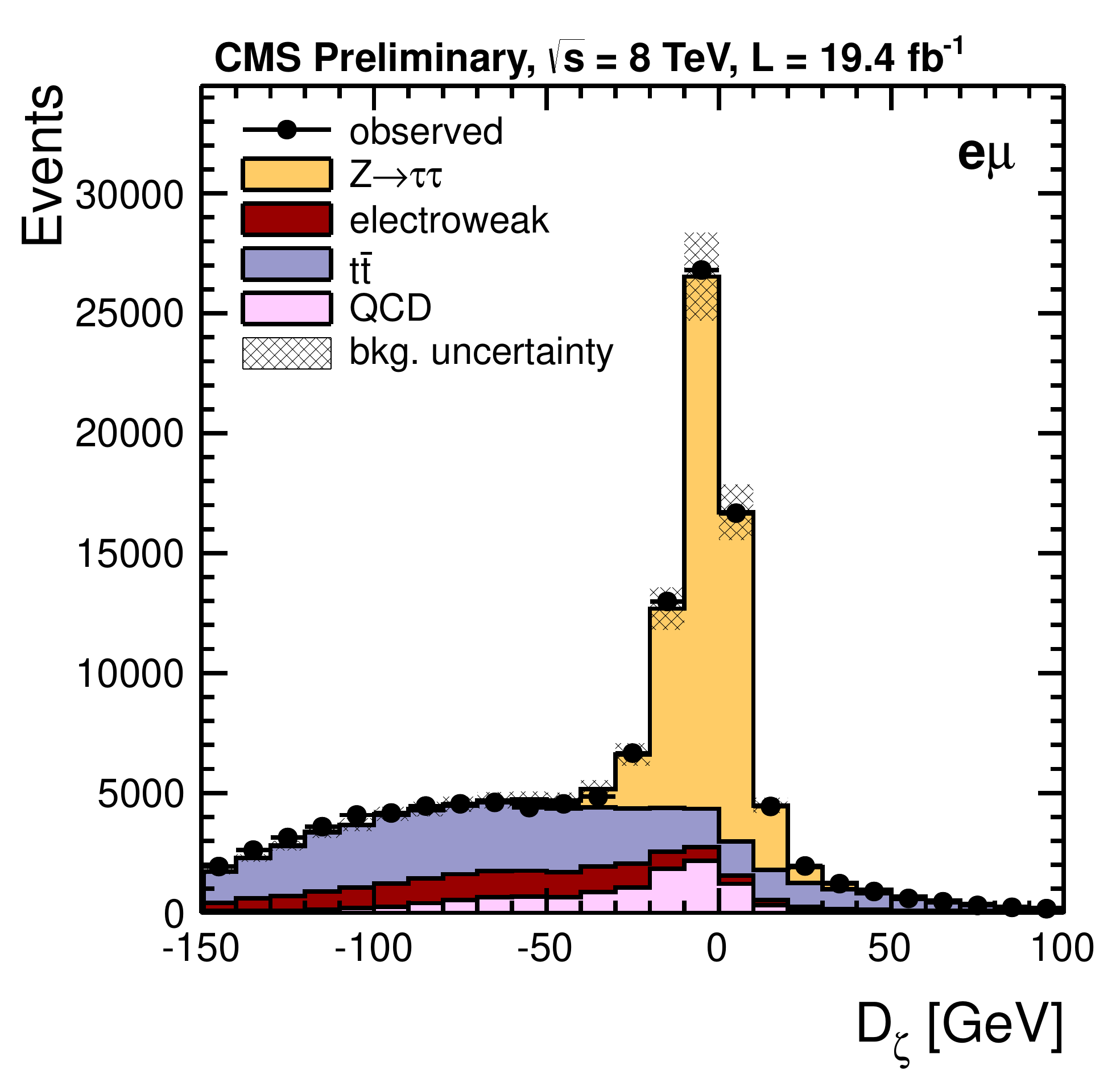}
\hspace{0.1cm}
\includegraphics[width=4.7cm]{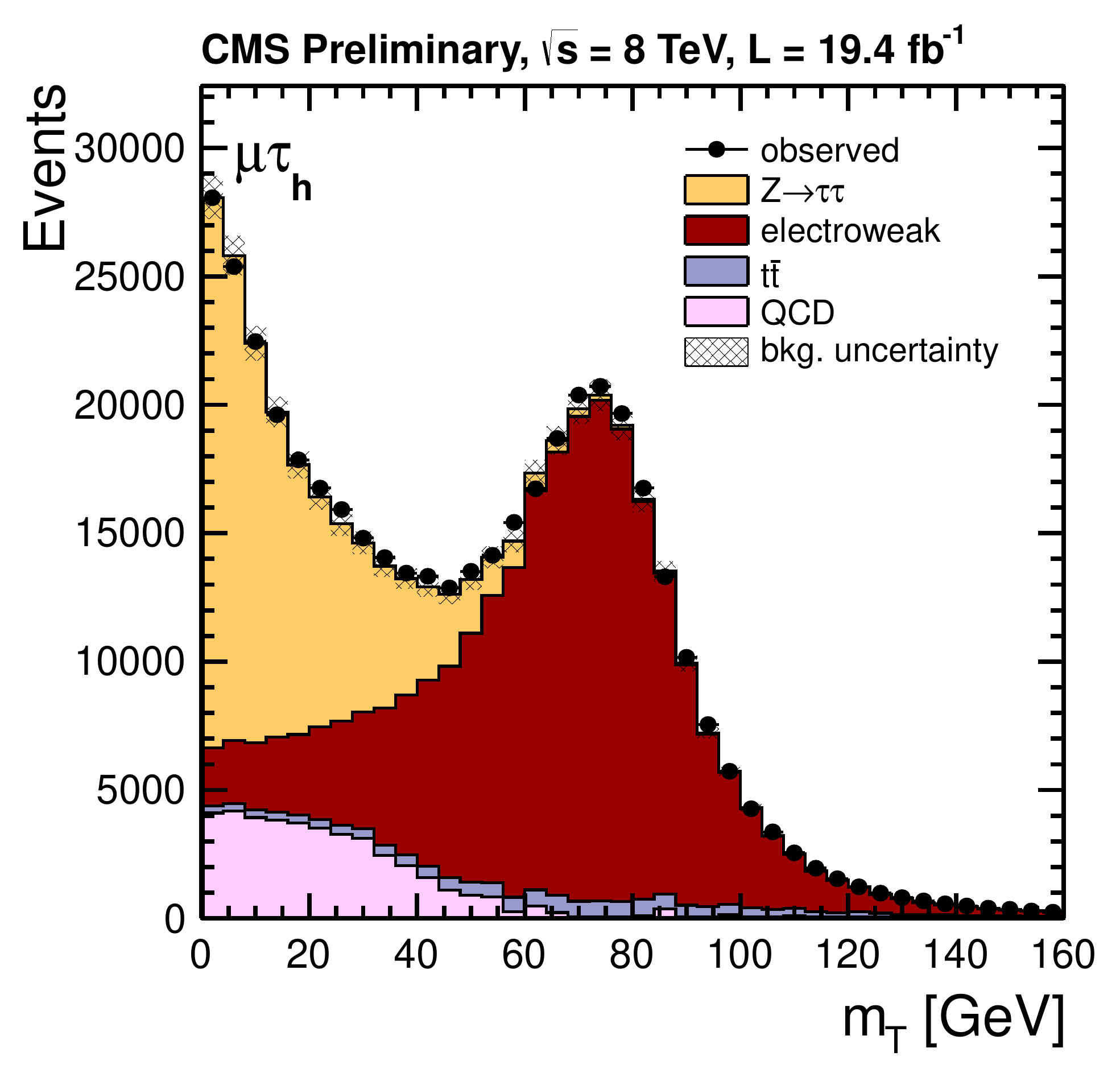}
\hspace{0.1cm}
\includegraphics[width=4.7cm]{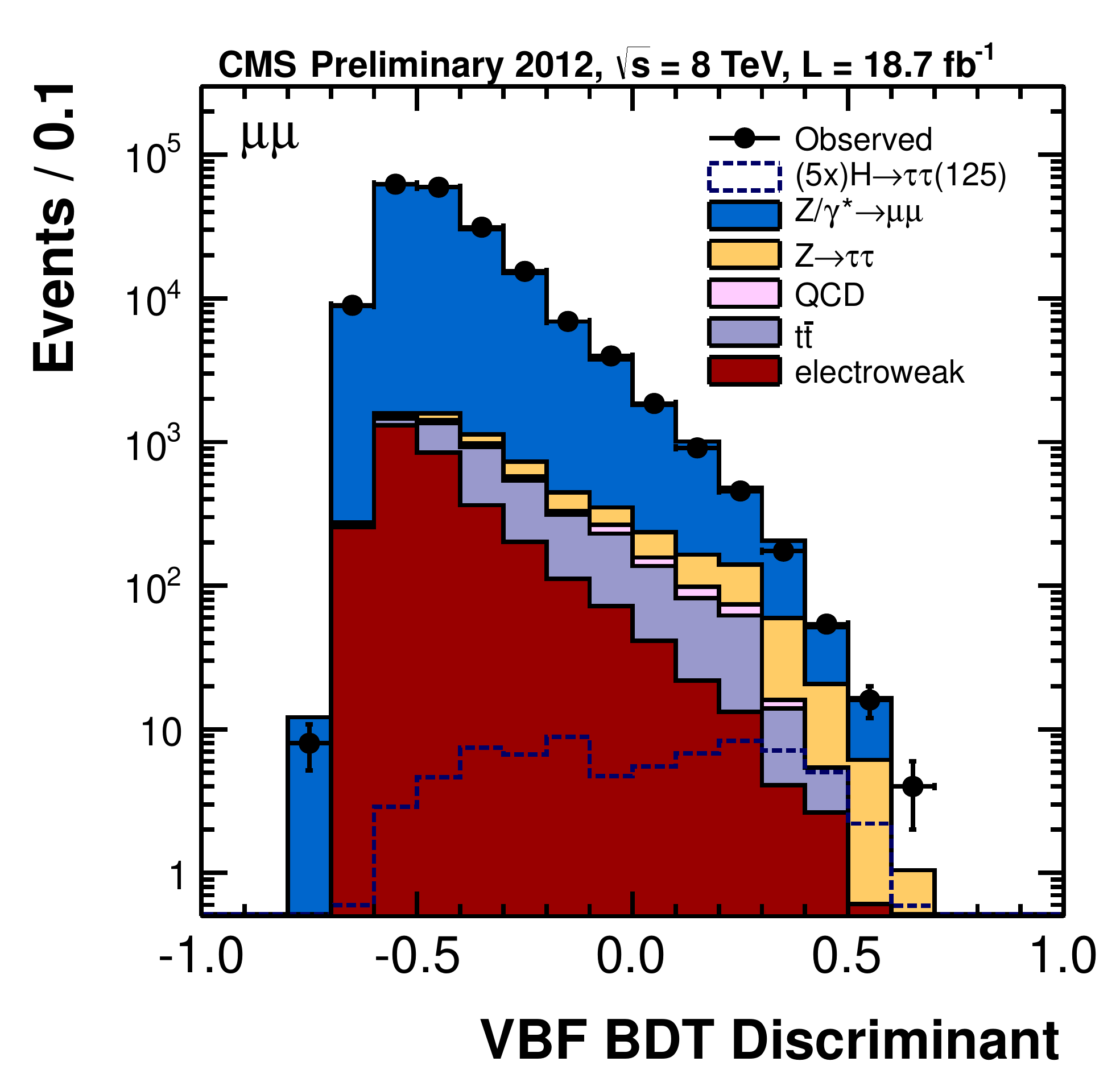}}
\vspace*{8pt}
\caption{Left: Distribution of the $P_{\zeta}$ variable in the $e\mu$ channel. Middle: Distribution of the $m_{T}$ variable in the $\mu\tau_{h}$ 
channel. Right: Distribution of the BDT discriminant in $\mu\mu$ channel as an example in the VBF category. \label{f2}}
\end{figure}

The background compositions are channel and category dependent with the $Z\gamma*\rightarrow \tau\tau$ processes representing typically the most dominant 
contributions over $t\bar{t}$, W+Jets, $Z\rightarrow\ell\ell$, QCD multijet-events and di-boson productions. To further 
reduce the various backgrounds additional topological event selection criteria depend on the final state are defined.

For the $e\mu$ final state the $t\bar{t}$ background is significantly reduced with selection criteria on $D_\zeta = p_\zeta - 0.85 \cdot p^{vis}_\zeta > -20 \textrm{ GeV/c}$, 
where $\zeta$ is the bisector of
the two leptons, $p_\zeta = \vec{p}_{T,1} \cdot \hat{\zeta}+ \vec{p}_{T,2} \cdot \hat{\zeta}+ \vec{E}^{miss}_{T} \cdot \hat{\zeta}$ and 
$p^{vis}_\zeta = \vec{p}_{T,1} \cdot \hat{\zeta}+ \vec{p}_{T,2} \cdot \hat{\zeta}$. $D_{\zeta}$ is a measure of how collinear the missing transverse energy 
vector is with the ransverse momentum of the di-lepton system (see figure~\ref{f2} (left)).

For the $e\tau_{h}$ and $\mu\tau_{h}$ final states the W+Jets background is reduced by selection criteria on the transverse mass variable 
$m_T = \sqrt{2p_T E^{miss}_T (1-cos(\Delta\phi))}$ (see figure~\ref{f2} (middle)).

For the $\mu\mu$ final state the overwhelming $Z/\gamma*\rightarrow \mu\mu$ background is reduced by selection criteria on dedicated multivariate Boosted Decision Trees
in each category.
(see figure~\ref{f2} (right)).

\section{Results}
The  observed and expected 95\% CL exclusion limit on the Standard Model Higgs Boson production rate is derived by a combined fit on the reconstructed di-tau mass $m_{\tau\tau}$ in
each category for each channel and shown in figure~\ref{f3} (right). For the $\mu\mu$ final state the 2 dimensional distribution of $m_{\tau\tau}$ versus 
the invariant mass of the two muons is used. Figure~\ref{f3} (left) shows the combined $m_{\tau\tau}$ distribution obtained after summing up the $m_{\tau\tau}$ 
distribution of each category of the $\tau_{h}\tau_{h}$, $e\tau_{h}$, $\mu\tau_{h}$ and $e\mu$ final states weighted by the ratio of the expected signal 
and background yields.  

\begin{figure}[h]
% \centerline{\includegraphics[width=4.7cm]{cmb_limit.pdf}
% \hspace{0.1cm}
\centerline{\includegraphics[width=4.7cm]{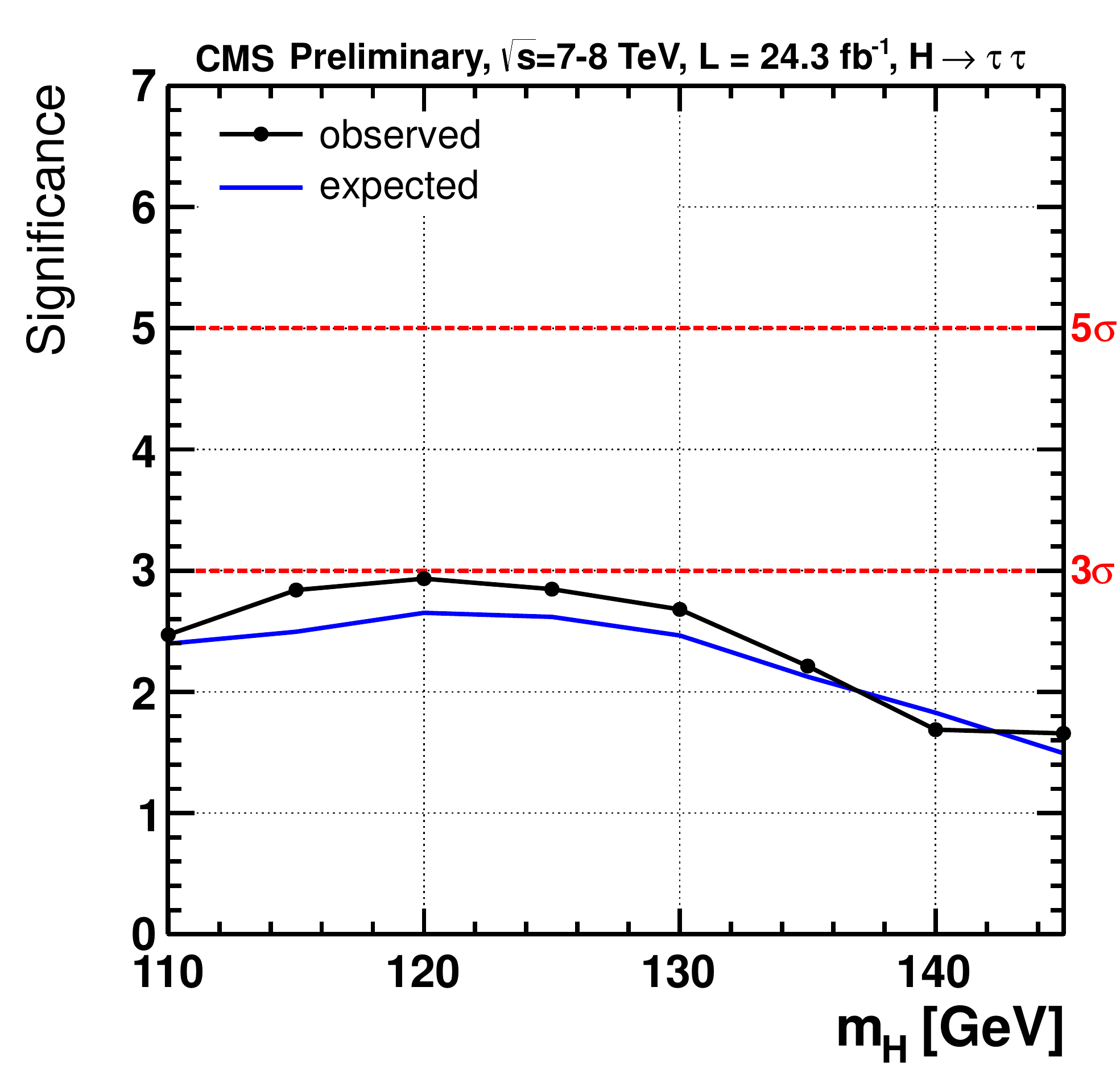}
\hspace{0.1cm}
\includegraphics[width=4.7cm]{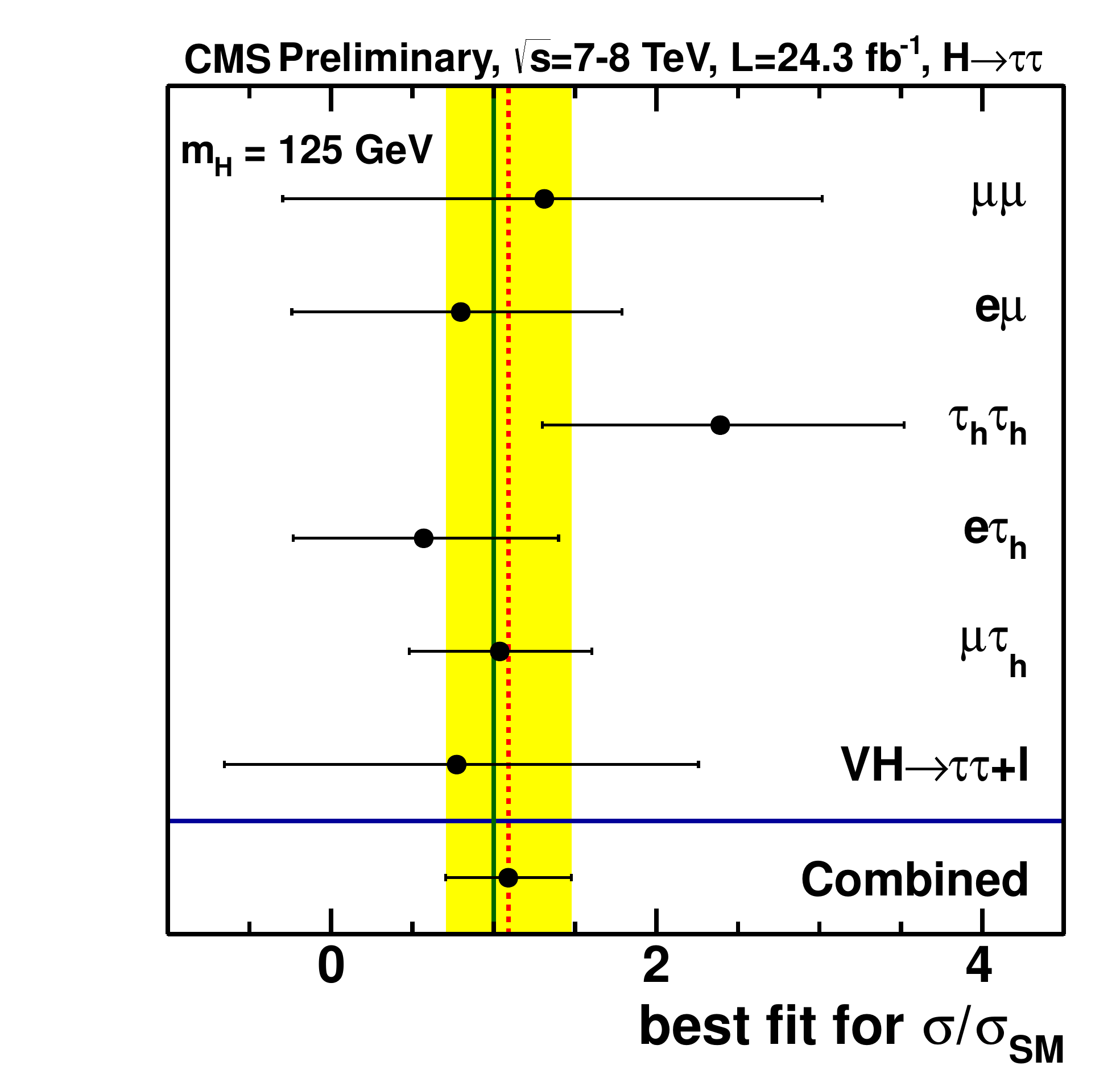}}
% \hspace{0.1cm}
% \includegraphics[width=4.7cm]{per_channel_extended.pdf}}
\vspace*{8pt}
\caption{ Observed and expected p-values as a function of $m_H$. Right: Best-fit signal strength at $m_H=125 \textrm{ GeV/c}$ relative to the standard model expectation.\label{f1}}
\end{figure}

An excess of the observed limit when compared to the background-only hypothesis is found and the local p-value and
signifiacance of this excess as a function of the Higgs mass is shown in figure~\ref{f1} (left). A maximum significance of 2.93 standard deviations 
is observed for a Higgs mass of $m_{H}=120 \textrm{ GeV/c}^{2}$ and a significance of 2.85$\sigma$ at $m_{H}=125 \textrm{ GeV/c}^{2}$ compared
to the Standard Model expectation of 2.63$\sigma$. The signal strength for $m_{H}$ is measured to be $1.1\pm 0.4$ times the Standard Model 
Higgs production rate (see figure~\ref{f1} (right))and thus the excess is compatible with the hypothesis of the presence of the Standard Model Higgs Boson.

\begin{thebibliography}{0}    %for 1 digit


\bibitem{TauPas} CMS Collaboration, {\it Search for the Standard-Model Higgs boson decaying to tau pairs in proton-proton collisions at sqrt(s) = 7 and 8 TeV}, 
CMS-PAS-HIG-13-004, 2-13

\bibitem{CMSHiggs} CMS Collaboration, {\it Observation of a new boson at a mass of 125 GeV with the CMS experiment at
the LHC}, Phys.Lett. B716 (2012) 30-61, doi:10.1016/j.physletb.2012.08.021, arXiv:1207.7235

\bibitem{AtlasHiggs} Atlas Collaboration,{\it Observation of a new particle in the search for the Standard Model 
Higgs boson with the ATLAS detector at the LHC }, Phys.Lett. B716 (2012) 1-29, doi:10.1016/j.physletb.2012.08.020, arXiv:1207.7214

\bibitem{CMSHiggsSM} CMS Collaboration, {\it Combination of standard model Higgs boson searches and measure-
ments of the properties of the new boson with a mass near 125 GeV}, CMS-PAS-HIG-13-
005, 2013.

\bibitem{CMSTauAssoPas} CMS Collaboration, {\it Search for the standard model Higgs boson decaying to tau pairs
produced in association with a W or Z boson}, CMS-PAS-HIG-12-053, 2013.
% @techreport{CMS-PAS-HIG-13-004,
%       title         = "{Search for the Standard-Model Higgs boson decaying to tau
%                        pairs in proton-proton collisions at sqrt(s) = 7 and 8
%                        TeV}",
%       institution   = "CERN",
%       collaboration = "CMS Collaboration",
%       address       = "Geneva",
%       number        = "CMS-PAS-HIG-13-004",
%       year          = "2013",
% }

% %%normal book (authors)
% \bibitem{autbk} R. Loren and D. B. Benson, {\it Introduction to String
% Field Theory}, 2nd edn. (Springer-Verlag, New York, 1999).
% 
% %%normal book (editors)
% \bibitem{edbk} R. Loren and D. B. Benson (eds.), {\it Introduction to
% String Field Theory}, 2nd edn. (Springer-Verlag, New York, 1999).
% 
% %%review volume
% \bibitem{rvo} C. M. Wang, J. N. Reddy and K. H. Lee, New set of
% buckling parameters, in {\it Shear Deformable Beams}, ed.~T. Rex
% (Elsevier, Oxford, 2000), p.~201.
% 
% %%book in a series
% \bibitem{seri} R. Loren, J. Li and D. B. Benson, Deterministic flow-chart
% interpretations, in {\it Introduction to String Field Theory},
% Ad. Series in Math. Phys., Vol.~3 (Springer-Verlag, New York, 1999),
% p.~401.
% 
% %%proceedings
% \bibitem{pro} R. Loren, J. Li and D. B. Benson, Deterministic
% flow-chart interpretations, in {\it Proc. 3rd Int. Conf.
% Entity-Relationship Approach}, eds. C. G. Davis and R. T. Yeh
% (North-Holland, Amsterdam, 1983), p.~421.
% 
% %%to be published
% \bibitem{publ} R. Loren, J. Li and D. B. Benson, Deterministic
% flow-chart interpretations, to appear in {\it J. Comput. System Sci.}
\end{thebibliography}
\end{document}